# Wear in multiple network elastomers arises from the continuous accumulation of molecular damage rather than microcrack growth

Ombeline Taisne[1,2], Julien Caillard[2], Côme Thillaye du Boullay[1], Marc Couty[2], Costantino Creton[1], Jean Comtet[1*]

[1]*Soft Matter Sciences and Engineering, ESPCI Paris, PSL University, CNRS, Sorbonne Université, 75005 Paris, France*
[2]*Manufacture Française des Pneumatiques Michelin, 63000 Clermont-Ferrand, France*

jean.comtet@espci.fr

**Tire wear releases millions of tons of particles annually, bearing immense industrial and environmental impact. However, efforts to mitigate the wear of elastomeric materials remain largely empirical, due to a limited understanding of the underlying damage mechanisms that cause wear. Employing mechanochemical approaches on model multiple network elastomers, we uncover how polymer strand scission events - the elemental damage units in these disordered networks - evolve during frictional wear. Our findings demonstrate that discrete micro-slippage at rough contacting asperities damages the material several micrometers below the surface through stress-activated bond scission events. This steady accumulation of subsurface damage ultimately leads to material erosion through the generation of a degraded viscous layer, painting a new picture of wear as a continuous damage growth process. Our approach further demonstrates an enhanced resilience to wear when tuning material architecture to reduce sensitivity to stress fluctuations, paving the way for knowledge-based strategies to develop more sustainable materials.**

## Introduction.

When two surfaces slide past each other, their relative motion is resisted by frictional forces. The science of friction has an extremely rich history, transitioning over several decades from a primarily engineering-focused field to a discipline grounded in solid physical principles. In most commonly encountered situations, friction is associated with wear, leading to the progressive degradation and restructuring of the contacting surfaces. The practical importance of wear for engineering and environmental concerns cannot be overstated: wear significantly impacts material lifespan, and material loss generates particles with increasing environmental implications. Yet this complex problem resists physical understanding and remains accordingly a blind (yet huge) spot in the field of tribology (1).

The challenges associated to the investigation of wear are manifold. First, the frictional interface is inherently buried between the two contacting bodies, composed of a myriad of interacting asperities that result in an emergent macroscopic response (2), (3). Second, wear involves complex damage processes, including coupled mechanical and chemical degradation, intrinsically difficult to disentangle and model (1), (4). Finally, the degradation of the surface of the material leads to the formation of an interfacial "third body", which affects the subsequent frictional and wear properties of the interface (5)–(7). These factors make wear an apparently intractable process.

In recent years, novel approaches driven by experimental or computational advances have offered new insights into this phenomenon, moving beyond traditional engineering and empirical methods. On the experimental front, studies at the scale of single-asperity contacts (4), (6), (8)–(12), potentially combined with atomic-scale imaging of single-asperity wear (12), (13), have illuminated the elementary processes involved in the wear of crystalline and metallic materials. These studies have revealed the occurrence of purely interfacial attrition processes, interpreted within the framework of stress-activated chemical kinetics (12) and elementary plastic deformation (13). On the computational side, the development of mesoscale simulations have provided new microscopic perspectives of the interplay between adhesion and crack propagation in the context of adhesive wear (14)–(16).

While these innovative approaches are well-suited to elastoplastic materials with a well-defined yield stress, the specific case of the wear of soft rubbery materials presents a distinct challenge. Since these materials do not locally yield before failure, identifying, characterizing, and modeling precursor damage events is highly challenging. Additionally, their deformable nature leads to peculiar behaviors at the level of contacting asperities, with potentially long-ranged mechanical stresses (17). Despite these complexities, elastomer wear is a highly pressing issue from both engineering and material perspective due to its immense importance for the transport industry. Tire wear alone is associated with an annual dispersion in the environment of several millions of tons of material (18).

The study of elastomer wear has a long history in its own right (19)–(23). Under typical conditions associated with smooth frictional loads, elastomer frictional wear exhibits a rich phenomenology. Surface morphogenesis manifests in various forms, such as wear ridges perpendicular to the sliding direction (24)–(28), rolls and particles (22), (29), (30) or smearing, i.e. the formation of a tacky degraded layer (19), (26), (31)–(33). These processes can coexist or evolve as the test progresses. The presence of these interfacial third bodies, generated by the wear process itself (5), (25), and potentially coupled with exogeneous mineral particles (34), can significantly impact subsequent wear formation. Due to the highly viscoelastic nature of rubbery materials, test conditions, such as sliding velocity and temperature also greatly influence wear rates (35), (36).

On the theoretical side, most approaches modeling fatigue wear consider crack growth mechanisms under steady-state or fatigue loadings to account for the formation of wear particles (37), (38), (39), (40). While these approaches provide good correlations between predicted and experimental wear particle size for large wear debris (37), they assume the pre-existence of micro-cracks in the material (41), (42) and do not account for smaller types of wear debris such as smear. Only a few approaches currently account for wear as a bulk fatigue process (43), (44).

These long-standing challenges motivate the development of novel strategies to achieve a more complete mechanistic understanding of the complex issue of soft elastomeric wear. In this work, we leverage innovative methods that combine mechanosensitive molecules (45)–(47) with quantitative imaging to detect and quantify molecular damage in the material resulting from sliding friction. We focus on a specific class of material, known as multiple network elastomers, providing here good model systems with stress-strain and toughness properties resembling particle-reinforced elastomers (48). By revealing the role of previously invisible molecular damage, our approach provides several new insights into elastomeric wear, which should stimulate further modeling efforts and guide the development of more resilient materials.

## Results and Discussion.

***Macroscopic frictional wear of multiple network elastomers in a multi-asperity contact regime.*** The principle of our tribological wear test is schematically shown in Fig. 1A, whereby a rough spherical glass bead of radius 5.2 mm is slid at a low velocity $v \approx 2$ mm.s$^{-1}$ in a linear back and forth motion on the elastomer surface, leading to the progressive abrasion of the material. The normal force $F_{\mathrm{N}}$ is adjusted through a dead weight and the resulting frictional force $F_{\mathrm{T}}$ is recorded simultaneously. The associated velocity and frictional signals are represented in the inset of Fig. 1A (see SI S2 and Materials and Methods for experimental details). The indenter is roughened with sandpaper, leading to a Root-Mean-Squared roughness of ≈ 1 μm over a 0.5x0.5 $mm^2$ area (Fig. 1B, see SI S3 for the full roughness spectral characterization).

Our materials of interest consist in a general class of reinforced elastomeric networks, known as multiple network elastomers (46), (48), bearing similarities with reinforced double network gels (49), (50). As pictured in Fig. 1D, these elastomeric networks have a nanocomposite architecture, with a fragile and stretched "filler" network (blue), interpenetrated with a highly deformable and loosely cross-linked matrix network (red). The degree of prestretch $\lambda_0$ [−] of the filler network can be tuned to vary material properties through successive swelling and polymerization steps. We will use here "double network (DN)" and "triple network (TN)" elastomers respectively, associated with a filler prestretch $\lambda_0 = 1.5$ and 2.5 and respective Young modulus of 1.3 and 1.7 (see SI S1and Materials and Methods). We focus in the first part of this paper mostly on DN elastomers, associated with a moderate prestretch of the filler.

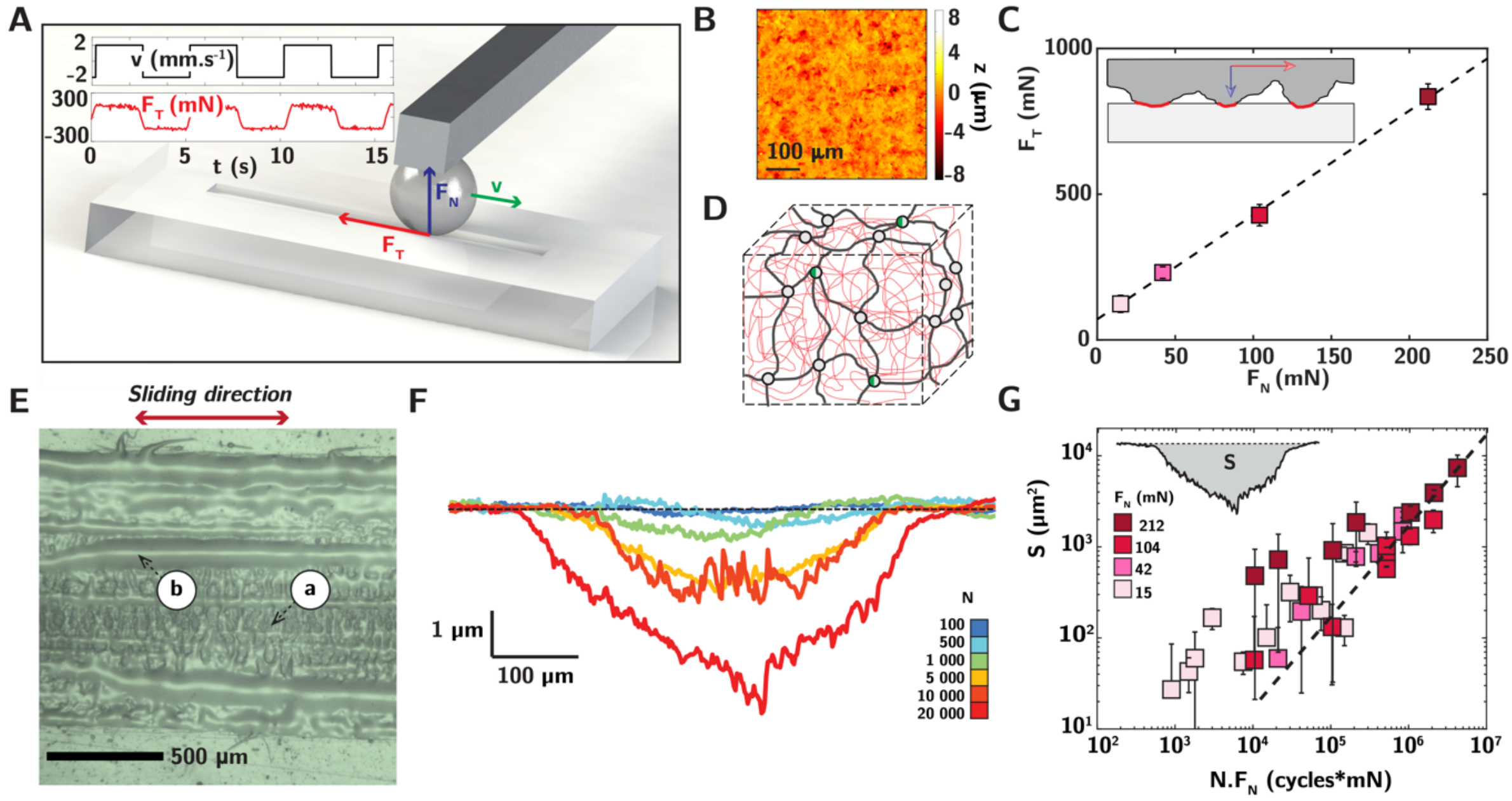

**Figure 1. Friction and wear of multiple network elastomers at the macroscopic scale. (A)** Schematic of the frictional wear test, whereby a roughened indenter slides periodically with relative velocity $v$ on the elastomer surface, with a normal loading force $F_N$, leading to progressive damage and wear of the material. The inset shows the alternating velocity profile and the associated friction force $F_T$ for a normal force $F_N = 42$ mN on a double network elastomer (see panel C). **(B)** Surface topography of the roughened indenter. **(C)** Evolution of the friction force $F_T$ with normal force $F_N$ for double network elastomer. The dashed line is a linear fit $F_T \sim \mu F_N$ with $\mu \approx 4$. The inset pictures the multi-asperity nature of the sliding interface. **(D)** Reinforced multiple network architecture with an interpenetrated structure between the filler network (blue) and the matrix (red). The prestretch $\lambda_0$ [-] of the filler network can be tuned during synthesis, with $\lambda_0 = 1.5$ and 2.5 for respectively double network (DN) elastomers and triple network (TN) elastomers. The mechanophore molecules (green/grey circles) are integrated at the crosslinking points in the filler network (see SI.S1). **(E)** Representative bright field picture of a reinforced TN elastomer after a high cycle number $N = 8000$ and $F_N = 42$ mN, evidencing (a) ridges perpendicular to the sliding direction, and (b) the generation of a liquid-like degraded third body. **(F)** Profilometric profile for DN elastomer at $F_N = 42$ mN and increasing friction cycle number $N = [100, 500, 10^3, 5 \cdot 10^3, 10^4, 2.10^4]$ (blue to red). **(G)** Evolution of the worn surface area $S$ as a function of $N \cdot F_N$, product of the cycle number by the normal force, with $F_N \in [15, 42, 104, 212]$ mN (pink to dark red). See SI.S6 for details and comparison with TN elastomer.

As shown in Fig. 1C, we observe a linear relation between the average friction force $F_T$ and the normal force $F_N$, following the phenomenological Amontons law, $F_T = \mu * (F_N + F_{adh})$, with $\mu \approx 3.6$ the effective friction coefficient and $F_{adh} \approx 20$ mN a residual adhesion force. This linear scaling between $F_T$ and $F_N$ is *a priori* not evident on such deformable materials and can be interpreted as stemming from two combined effects. The roughness of the indenter leads to a discontinuous contact at the microscopic scale (51), (52), with a real area of contact $A_\mathrm{R}$ associated with protruding asperities of the indenter accounting for 10 to 20% of the apparent contact area and increasing linearly with the normal load as $A_\mathrm{R} \sim F_\mathrm{N}$ (see SI S3). The Amontons law is then recovered by assuming that the frictional force scales with this real contact area and can be expressed as $F_\mathrm{T} = A_\mathrm{R}.\tau_0$, where $\tau_0 \approx 2$ MPa is the characteristic shear stress accounting for local dissipation between the contacting asperities. This shear stress is here expected to have from both viscoelastic and adhesive contributions (see SI S4), consistent with the large friction coefficient $\mu > 1$.

Sliding the indenter on the surface under reciprocal motion, we observe the progressive wear of the elastomeric material, associated with surface restructuring. As shown for a TN elastomer in Fig. 1E, the surface shows typical features of rubber wear (6,7), with the formation of wear ridges perpendicular to the sliding direction (arrow *i*). Concomitant to these morphological changes, frictional wear leads also to the formation of a third body, which takes here the form of a gooey liquid, known as smear (19), resulting from the microscopic debris detached from the elastomer surface (arrow, *ii*). We note that in real-world conditions, such smearing layer can mix with exogenous minerals, leading to the formation of particles (20), (34), (53).

We turn to Fig. 1F for the quantitative evaluation of the macroscopic wear process. The amount of worn material is assessed through the quantification of the abraded area $S$ (m$^2$) below the pristine surface by profilometry (SI S5 and Materials and Methods). This quantity is directly proportional to the abraded volume $V$ [m$^3$], with $V = S.L$ where $L$ is the sliding length over one cycle, classically estimated through gravimetric approaches (35), (54). The profiles in Fig. 1F correspond to a normal force $F_\mathrm{N} = 42$ mN and show a steady increase of the volume of abraded material with the number of sliding cycles (blue to red). At the maximal number of $10^4$ sliding cycles, the wear marks reach depths of approximately 10 µm for a width of about 0.5 mm.

At a phenomenological level, we expect material wear to occur at the contacting asperities between the indenter and the surface. In this situation, the worn volume $V$ can be simply expressed as an *Archard law* (55), with $V = k.A_\mathrm{R}.l$, where $A_\mathrm{R}.l$ is the equivalent "volume of interface" seen by the material, written as the product of the real area of contact $A_\mathrm{R}$ by the total sled distance $l = N \cdot L$. Assuming a linearity between $A_\mathrm{R}$ and $F_\mathrm{N}$ (Fig. 1D), this equality amounts to $S \sim F_\mathrm{N}.N$. As shown in Fig. 1G and SI. S6, such relation compares well with our experimental data. The dimensionless factor $k$ comparing worn and sled volumes can be understood here as an elementary material removal probability. It is found of the order of $10^{-6}$ for the DN elastomer, meaning that one per million strand wears off following contact. Note that this Archard-like approach is equivalent to an approach based on a mechanical energy balance, bearing also a linearity between the worn volume $V$ and the frictional forces $W = F_\mathrm{T}.l$ dissipated in the contact, with $V \approx k/\tau_0 \mathrm{W}$.

***Damage sensitive mechanochemical probes reveal subsurface damage by chain scission following frictional sliding.*** To go beyond these macroscopic approaches, we turn in Fig. 2 to the evaluation of molecular damage by chain scission in the network following frictional wear. The principle of our mechanosensitive approach is presented in Fig. 2A, whereby we label the filler network of our multiple network elastomers with a mechanophore cross-linker, the Diels-Alder Cross-Linker (DACL) (see materials synthesis SI.S1 and Materials and Methods). Being incorporated as a cross-linker, our DACL mechanophore is sensitive to molecular-scale forces applied to this force-bearing network (46). These approaches have provided recent insights into material damage by fracture (45)–(47), fatigue (56), or more complex mechanical processes such as cutting (57) and cavitation (58).

As shown in Fig. 2A.a, the mechanophore is non-fluorescent in its native form. However, if the connecting strands are under sufficient tension the Diels-Alder adduct bond of the mechanophore can "break" at a slightly weaker force than a C-C bond, or in more chemical terms, undergo a force activated retro-diels-alder reaction which will lead to the release of a fluorescent pi-extended anthracene moiety (Fig. 2A.b). This retro-Diels-Alder reaction is irreversible at room temperature. Combining high quantum yield (0.72), resistance to oxygen quenching and to visible light excitation (59), this mechanophore constitutes an excellent reporting molecule for strand scission and allows its post-mortem detection using subsequent fluorescence mapping. In particular and as previously described (46), (47), the mechanophores behave as slightly weaker points in the force chain, allowing the evaluation of the fraction of broken crosslinkers $\phi$ based on the measurement of the concentration of activated mechanophore $c_{\text{activated}}$ and concentration of incorporated mechanophore $c_0$ as $\phi = c_{\text{activated}}/c_0$ ((60) and SI.S8).

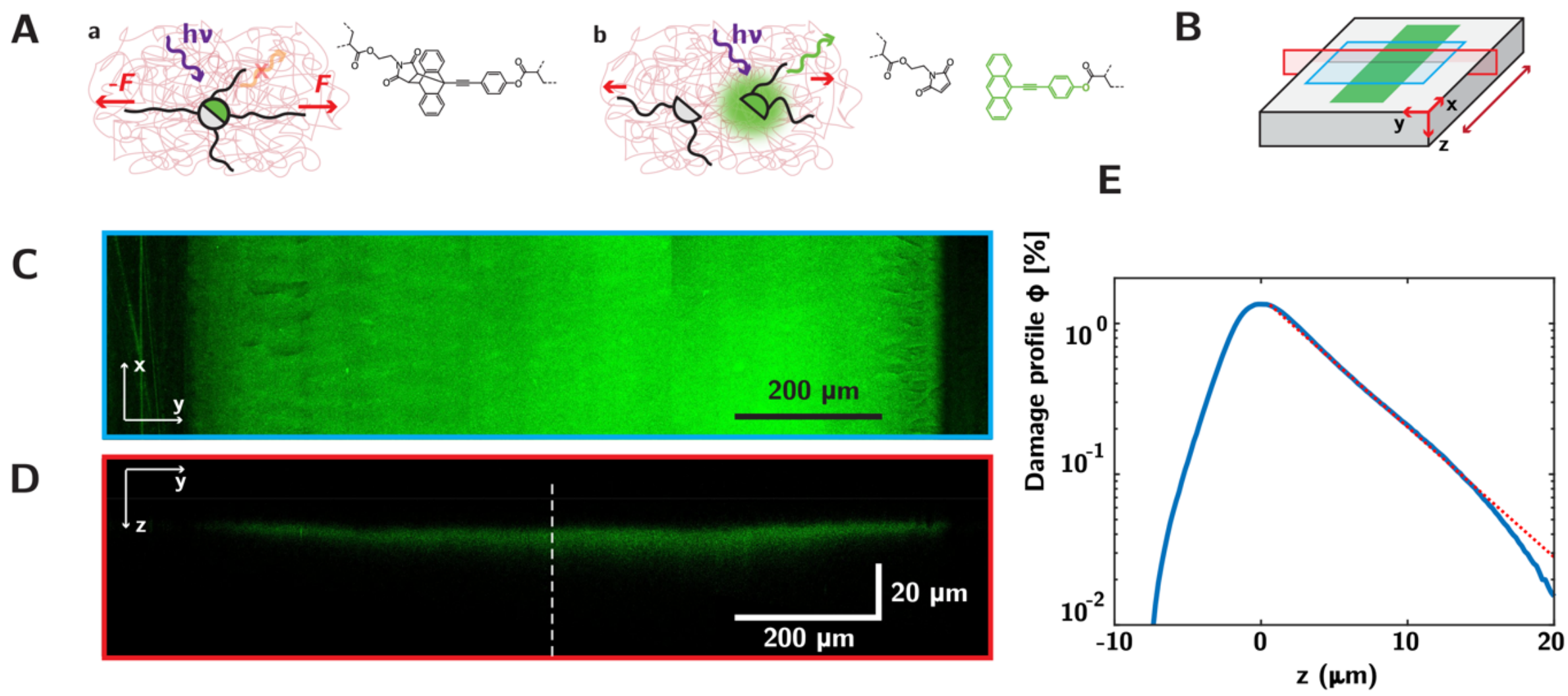

**Figure 2. Mechanophores as molecular damage reporters following frictional wear. (A)** Mechanophore activation reports for damage by chain scission in the elastomeric material. The filler network is represented in black and the matrix in red. Mechanophore are included as crosslinking points in the filler network. **(a)** Non-fluorescent form of the mechanophore, connected to a strand under tension. **(b)** Irreversible scission of the mechanophore, leading to the release of a fluorescent anthracene moiety reporting for network scission (in green). **(B)** Imaging planes normal (blue) and perpendicular (red) to the surface, with the wear mark in green, and the sliding direction indicated by the red arrow. **(C-D)** Representative fluorescence intensity maps parallel (C) and perpendicular (D) to the surface, obtained on a double network, at 1000 cycles and $F_N = 212$ mN. **(E)** Damage profile $\phi(z)$ associated with the fraction of broken chains in the depth of the material, evaluated over a 270x270 $\mu m^2$ surface area. The red line is an exponential fit.

As shown in Figs. 2B-D, we thus turn to confocal microscopy to map the spatial distribution of mechanophore activation in the material after a large number of friction cycles (here $N = 1000$ cycles at $F_N = 212$ mN on DN elastomers). The obtained 3D volumetric scans of the fluorescence intensity are projected respectively parallel and perpendicular to the elastomer surface (Fig. 2B, see Materials and Methods for imaging conditions). Considering first the projection of the fluorescence intensity perpendicular to the sliding surface (Fig. 2C), the wear mark stands out clearly in these fluorescent maps, demonstrating the occurrence of large damage levels localized on the region of contact with the sliding indenter. The 3D nature of our confocal microscopy mapping allows us to also probe how far the activation extends below the surface of the material (Fig. 2D). Remarkably, rather than a purely localized fluorescent intensity at the extremum of the elastomer surface, we observe that activation levels spread out on a relatively thick region of several tens of micrometers from the surface to the material bulk, demonstrating clearly the occurrence of spatially extended and diffuse damage gradients. Notably, no heterogeneous characteristic features such as crack precursors are present in these images.

In order to characterize this diffuse profile of mechanophore activation, we convert the local fluorescence intensity into a damage variable expressed as a fraction of broken chains $\phi$ (using calibration samples, see SI.S8). As represented in Fig. 2E, this spatial profile $\phi(\mathrm{z})$ extends over tens of micrometers from the surface and decays approximately exponentially in the material as $\phi \propto \phi_{\max} e^{-z/\lambda}$, with a characteristic decay length $\lambda \approx 2-7$ µm. The observation of such spatially extended damage suggests the presence of diffuse levels of stresses extending well beyond the elastomer extreme surface following sliding, a point we will address below. We highlight that the variable $z$ characterizes here the distance to the surface of the material, and that the occurrence of a non-zero fluorescence intensity for $z < 0$ stems from the finite numerical aperture of our imaging system (see SI S9).

***Damage accumulates in a spatially localized fashion through micro-slippage events.*** To gain more insights on the mechanisms leading to the diffuse extension of damage in the sub-surface of the material, we focus in Figure 3 on the way damage accumulates and evolves at a low number of cycles. This regime is traditionally inaccessible to macroscopic frictional wear measurements as it is not associated to any mass loss nor apparent changes of the surface. However, the high sensitivity conferred by our mechanochemical approach allows us to observe a clear activation of the mechanophore signal since damage precursor events occur even for the first friction cycle (see SI.S10 for comparison with pristine samples). To characterize this spatial distribution of molecular damage, we represent in Fig. 3A, the projected map of damage activation parallel (left) and perpendicular (right) to the surface, for an increasing number of friction cycles. These maps show a striking evolution in the spatial distribution of damage. After the first friction cycle, we observe highly heterogeneous fluorescence intensity maps, showing the presence of localized activation patches in the plane of the surface. Interestingly, these patches show a clear anisotropy, extending in the y-direction perpendicular to the sliding direction, echoing with observations of shear-induced contact anisotropy observed in soft macroscopic elastomeric contact (61)–(63), yet they are here occurring at the scale of the microscopic contact asperities. The in-plane spatial extension of these damage zones is of the order of $\approx 10$ µm and coincides with the typical size of contact asperities, set by the multi-asperity nature of the contact (SI.S3). Remarkably and as shown in the subsurface profiles (Fig. 3A, rectangular side panel), these in-plane spatial heterogeneities extend below the surface in the form of 3D localized intensity patches, with a typical depth similar in order of magnitude with the lateral contact extension.

Comparing the various images in Fig. 3A, we observe a clear homogenization of the spatial damage profile when increasing the number of sliding cycles, which we attribute to a cumulative effect, set by the successive passage of the sliding asperities on the elastomer surface. To quantify these surface heterogeneities, we compute in Fig. 3B the histograms of the projected surface damage, expressed as a local areal fraction of broken chains as

$$\sigma(x,y) = \nu_{\mathrm{x}} \int_0^\infty \phi(x,y,z)dz \qquad \textit{(Eq. 1)}$$

Where $\nu_{\mathrm{x}}$ is the crosslinking density (SI.S1 and SI.S11). To highlight the occurrence of spatial heterogeneities, this local areal fraction is calculated over elementary surface of size 758*758 $\mathrm{nm}^2$. This histogram confirms our qualitative observations: for a single cycle (dark blue), the distribution of damage at the surface is broad and peaks at low values. At intermediate number of cycles, increasing the number of sliding cycles shift the distribution to higher damage values while reducing its width. For 1000 cycles, the distribution is the narrowest, while we also observe a decrease of the mean surface damage, a surprising point to which we will come back below.

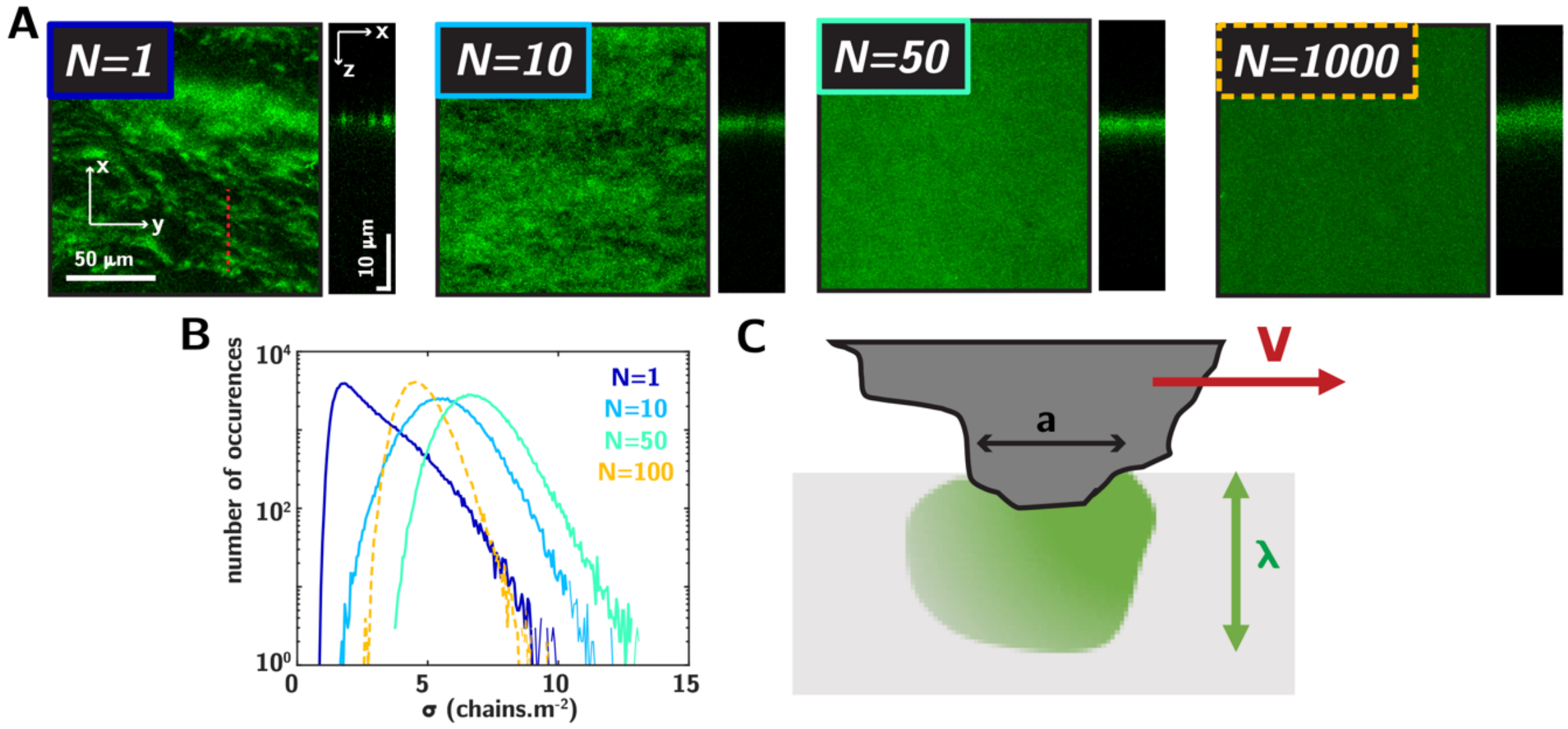

**Figure 3. Damage accumulates in a spatially heterogeneous fashion through micro-slippage events**. **(A)** Upper images show the maximum intensity projection at the wear mark surface for increasing cycle number ($N = 1, 10, 50, 1000$ for DN elastomer at $F_N = 42$ mN). Sliding is in the vertical x-direction, and the surface shows the presence of localized and heterogeneous activation patches. The contrast on each image was adjusted to highlight these local events. The associated vertical rectangular panels highlight reconstruction of the intensity below the surface, with a scale bar of 10 µm both in the vertical ($z$) and horizontal ($y$) directions. The first panel $N = 1$ is shown in duplicate in Fig. S10. **(B)** Quantification of damage heterogeneity, showing the distribution of areal damage $\sigma$ [chains.m$^{-2}$] for the four conditions in $A$, with respectively blue, cyan, azul and yellow associated with $N = 1, 10, 50$ and 1000 cycles. **(C)** Schematic representation of a micro-slippage event at the scale of one asperity, inducing a spatially extended stress and damage in the bulk of the material over a length $\lambda$ set by the contact patch area $a$ with $\lambda \approx a$.

At a qualitative level, classical contact mechanics approaches predict stress fluctuations to penetrate in the bulk of the material over length scales set by the local contact asperity size (64), (65). While this picture would be consistent with the observation of a damage field extending over micrometer distances (Fig. 2E), the spatially-resolved observations of discontinuous damage distribution in the plane of the surface suggests the occurrence of more complex microscopic processes at the interface. We propose here that such discontinuous accumulation of damage at the surface stems from loading events at the scale of the contacting asperities, associated with local stick-slip instabilities or detachment and reattachment fronts, leading to heterogeneous sliding motion. Such discontinuous damaging processes in multi-asperity contact, associated to heterogeneous motion at the scale of the contact asperities echoes toward recent microscopic observations of rough contact sliding (62).

***From damage accumulation to erosion.*** Following the observations of spatial heterogeneities in the damage accumulation process, we turn in Fig. 4 to the quantification of the total amount of accumulated damage following sliding, and its dependence on cycle number and normal force. The normal forces range here from 15 mN to 212 mN leading to macroscopic pressures from 0.4 to 1 bar. To characterize damage evolution, we focus here on the evolution of a spatially averaged integrated quantity $\bar{\Sigma}$ [-] corresponding to an effective number of broken monolayers and defined as:

$$\bar{\Sigma} = \langle \sigma \rangle_{x,y} / \Sigma_0 \qquad (Eq.\ 2)$$

Where $\langle\sigma\rangle_{x,y}$ is the number of broken chains averaged over an extended surface area of size 270*270 μm$^2$ taken in the center of the wear mark and larger than the typical spatial extension for damage fluctuation and $\Sigma_0$ correspond to the number of broken bonds needed to create two surfaces (effectively a "monolayer" in terms of network unit mesh, see SI.S11).

As shown in Fig. 4, $\bar{\Sigma}$ shows a clear trend with both the number $N$ of sliding cycles and the normal force. For relatively low cycle number, we observe for each of the normal forces, a monotonous increase of $\bar{\Sigma}$ with the number of sliding cycles $N$ (dashed lines, Fig. 4A). The rate of damage accumulation further shows an increase for increasing normal forces (going from red, green, yellow to brown). Upon further increase in the number of sliding cycles, we observe $\bar{\Sigma}$ to decrease and reach a steady-state value, which appears to be weakly dependent on the contact pressure or normal force.

The transition between these two damage regimes coincides approximately with the onset of material removal for the three lowest normal forces, which we highlight by the downward pointing colored arrows in Fig. 4A (yellow, green and red). We dub the first regime of steady damage growth at low cycle number as one of "damage accumulation", whereby molecular damage increases in the material due to the successive sliding cycles, without significant material loss (characterized in the literature as an incubation period (43)). The second regime is characterized as one of erosion, whereby a balance is established between accumulation of damage by the sliding indenter, and the subsequent removal of material.

Yet, we observe that the transition between these two regimes of damage seems to occur at a value $N \approx 1000$, independent of the normal force. There is accordingly not an exact correlation between this transition in damage evolution, and the onset of material erosion, as characterized by profilometry (Fig. S7 and colored arrows on Fig. 4A). In particular, for the highest load of 212 mN, the onset of material erosion (brown arrow, $N \approx 50$) appears significantly earlier than the drop of $\bar{\Sigma}$. Similarly, for the lowest normal load of 15 mN, the end of the steady damage growth regime (end of the linear extrapolation with red dotted line) appears slightly before any significant material erosion (red arrow). These discrepancies might appear surprising. Yet, we first stress that as shown in Fig. S7, there is some uncertainty in the determination of the exact value of the onset of erosion especially for the three lowest normal forces. Second, we remind that $\bar{\Sigma}$ is an integrated quantity, and that its evolution is set by a relative balance between erosion and damage accumulation, which shows a complex and non-linear dependency on $N$ (anticipating our results below). The exact onset for the drop of $\bar{\Sigma}$ might then not be directly correlated with the onset of material removal. A more refined evaluation of this transition could be obtained by probing the evolution of the spatially resolved damage profile below the surface, in comparison with the characteristic depth of the wear mark. We leave such refinements for further studies.

***Non-linear damage accumulation and the role of network disorder.*** We first focus on the evolution of damage in this first *accumulation* regime. As the removal of material from the elastomer surface is negligible, this initial regime allows to probe quantitatively how damage induced by the sliding indenter progressively accumulates following successive sliding cycles. A striking observation is the highly non-linear nature of the damage accumulation process, evidenced by the logarithmic fits (straight lines in Fig. 4A) with molecular damage increasing slowly and sub-linearly with the number of cycles, such that

$$\bar{\Sigma} \propto \log(N) \quad \textit{(Eq. 3)}$$

This sublinear growth suggests the occurrence of complex non-linear processes during damage accumulation, and is in clear contradiction with classical fatigue or damage accumulation models, such as the Miner rule of linear damage accumulation, widely used in mechanical engineering (66). Remarkably, the effect of contact pressure can be accounted for in Fig. 4B, where we observe a good collapse of our data in the accumulation regime when normalizing damage by the contact pressure $P$, leading to the phenomenological law:

$$\bar{\Sigma} \sim P/E \cdot \log(N) \qquad (Eq.\ 4)$$

with $E$ the Young modulus of the material.

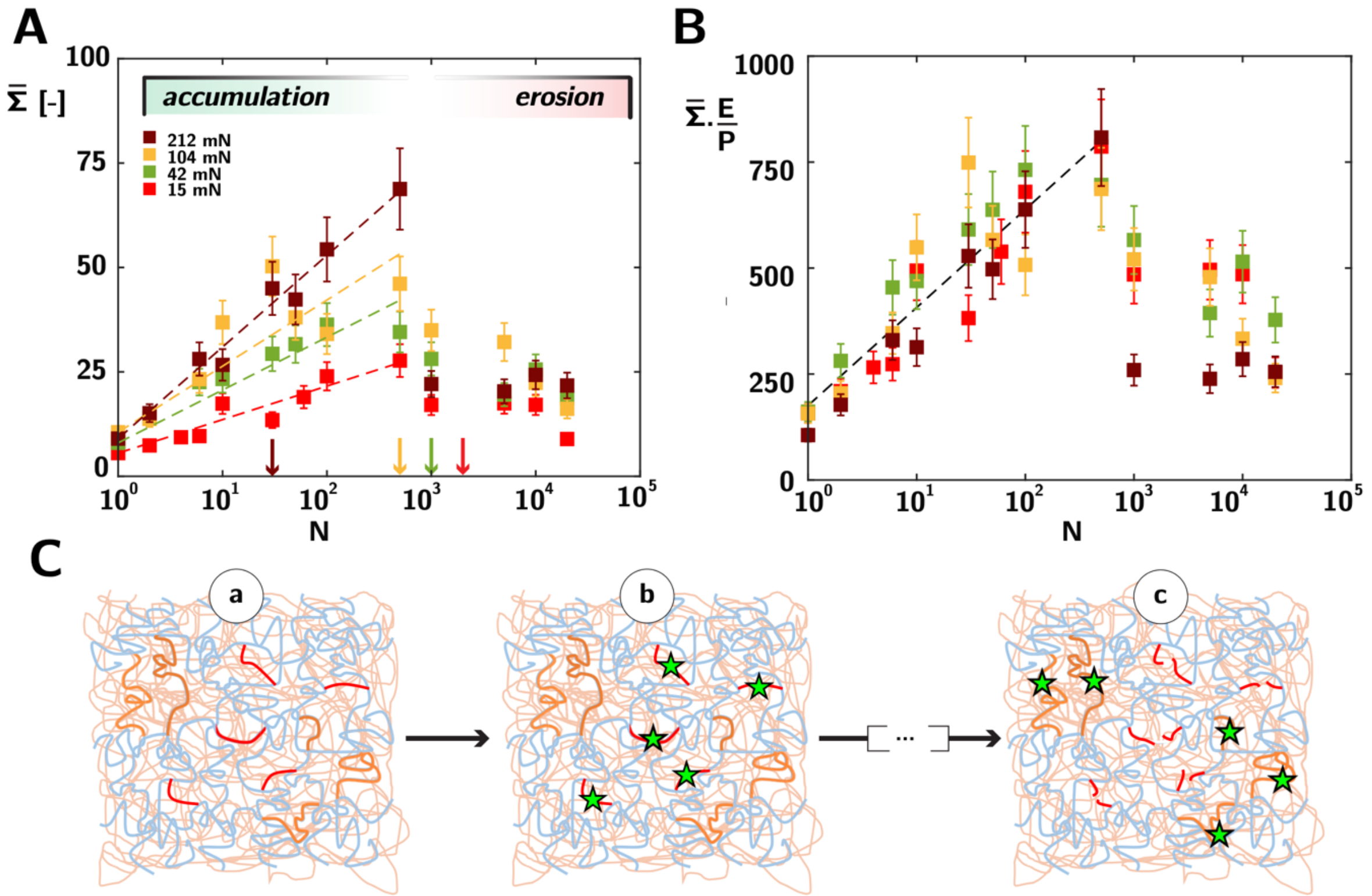

**Figure 4. Integrated evolution of damage in the accumulation and erosion regime. (A)** Evolution of molecular damage, $\bar{\Sigma}$ [-], as a function of the number of cycles for different normal forces of respectively 15, 42, 104 and 212 mN for red, green, yellow and brown symbols. The error bars are representative of the deviation of $\bar{\Sigma}$ on three wear marks obtained using the same conditions. The colored arrows indicate the onset of material erosion (see SI S7), setting the transition from accumulation to erosion regime. Dashed lines are logarithmic fits of the data up to 500 cycles with: $\bar{\Sigma} = \alpha \log(N) + \bar{\Sigma}_{N_{cycle}=1}$ and $\alpha =$ 2.3, 3.5, 4.4, 6.1 ($10^{17}$.chains.m$^{-2}$) for $F_N = 15, 42, 104, 212$ mN respectively. **(B)** Evolution of $\bar{\Sigma} \cdot E/P$ with $N$ showing the dependence of molecular damage on apparent contact pressure. The dashed line is a logarithmic fit of the data in the accumulation region. Coefficient of logarithmic fit is 100. **(C)** Slow damage accumulation in a heterogeneous network. **(a)** We depict the strands with a high stored elastic energy in red (stretched) and the ones with a lower elastic energy in orange (coiled). **(b)** The red strands break at low number of cycles (green stars). **(c)** A much larger number of loading cycles is required to break the orange chains with a slightly lower stored elastic energy.

Despite its peculiarity, this logarithmic growth of damage *with the number of friction cycles* is reminiscent of logarithmic aging growth law *with time* which can be observed in a large range of complex systems, e.g. shear strength of single sliding nanocontacts (67), avalanche angle in humid granular media (68), or the crumpling of paper (69). The common prevailing interpretation for such logarithmic aging law lies in the combination of thermally activated processes associated with a broad distribution of energy barriers due to disorder, with higher energy barriers being explored over increasing time, leading to a slow yet steady logarithmic growth.

Using this analogy, we consider the network strands under stress to be characterized by a wide distribution of elastic energies, and the passage of each asperity on the elastomer surface to represent an elementary attempt to break these chains. If we consider a chain associated with a given stored elastic energy $W$, its rate of scission can be expressed as a stress-activated scission event (70) as

$$k_{\mathrm{off}} = k_0 . \exp(W/k_{\mathrm{B}}T) \qquad \textit{(Eq. 5)}$$

with $k_0$ ($s^{-1}$) an equilibrium scission rate. Considering a *fatigue* perspective, the mean number of attempts $n(W)$ needed to break this chain will scale as $k_{\mathrm{off}}^{-1}$ and can be expressed as

$$n(W) \sim \frac{1}{\tau_{\mathrm{load}} * k_0} . \exp(-W/k_{\mathrm{B}}T) \qquad \textit{(Eq. 6)}$$

with $\tau_{\mathrm{load}}$ an elementary microscopic time spent under load during each microscopic fatigue events. This expression highlights the highly non-linear relation which arises between the energy stored in the elastic chain and the associated fatigue lifetime of the chain: chains with a high stored elastic energy will break very easily over the first few sliding cycles but it will then become increasingly difficult (taking an increasingly longer number of cycles) to break the following chains with only slightly lower $W$. Within this picture the distribution of elastic energies sets the damage evolution in the network, so that a broad energy distribution should lead to a slow increase of molecular damage with $N$.

Probing the microscopic state of the material following $N$ friction cycles, we expect from Eq. 6, scission of all chains with $W > W_{\mathrm{N}}$ where $W_{\mathrm{N}}$ decreases with the number of attempts as

$$W_{\mathrm{N}} \sim W_0 - k_{\mathrm{B}}T \cdot \log(N) \qquad \textit{(Eq. 7)}$$

with $W_0 = -k_{\mathrm{B}}T \cdot \log(\tau_{\mathrm{load}} k_0)$, a reference activation energy.

In order to obtain the evolution of the absolute number of broken chains with the number of cycles, we need to know how elastic energies are distributed within the network. For an arbitrary energy density distribution $\rho(W)$ the number $\mathcal{N}_{\mathrm{chain}}(N)$ of broken chains can be written as $\mathcal{N}_{\mathrm{chain}}(N) = \int_{W_{\mathrm{N}}}^{\infty} \rho(W) dW$. While the distribution for $\rho(W)$ is not known *a priori*, assuming a *locally* uniform density distribution for $\rho(W) \approx \rho_0$, consistent with disorder in the network, and a broad distribution for $W$ amounts to

$$\mathcal{N}_{\mathrm{chain}} \sim \rho_0 k_B T \cdot \log(N) \qquad \textit{(Eq. 8)}$$

(dropping the integration constant associated with the asymptotic behavior $\rho \to 0$ as $W \to \infty$). Despite its simplicity, our approach predicts the experimentally observed logarithmic increase of broken crosslinks shown in Fig. 4.

To delve further in this microscopic damage approach, we consider the effect of the normal force, accounted for as shown in Fig. 4B by the phenomenological relation $\bar{\Sigma} \sim P/E \cdot \log N$. Here, the multi-asperity nature of the contact suggests that an increase of the normal force causes an increase in the real area of contact $A_{\mathrm{R}}$, rather than an increase in the local pressure at these contact patches (71). Increasing the local contact pressure thus amounts to an increase in the fractional area of contact $A_{\mathrm{R}}/A$, with $P/E \sim A_{\mathrm{R}}/A$, leading to $\bar{\Sigma} \sim A_{\mathrm{R}}/A \cdot \log N$ (with $A$ the apparent area of contact). This expression highlights again the role of the contact pressure in setting the damage rate, due to an increase in the areal density of sliding asperities.

These non-linear effects in damage accumulation highlight the key role played by the material history in response to fatigue cycles, so that the assumption of each friction cycle being independent of previous ones is highly unrealistic. The broad distribution of stored elastic energy could stem from heterogeneities in chain length distributions combined with a complex redistribution of stresses at the molecular scale. While our experimental trend is well captured by our simplified mean-field approach, more complex phenomena could be considered and call for detailed modelling. First, we neglect in our mean-field description the larger-scale spatial heterogeneities associated with discontinuous damage accumulation (Fig. 3), which will lead some high-energy chains to remain unaffected even after the first sliding cycles. Properly accounting for these effects would require some homogenization process for the spatial damage, which we leave for future work. Second, our approach also neglects the role of the progressive stress redistribution in the network following damage of a chain subpopulation, which could have large consequences on the distribution of stored elastic energy in the network, and the subsequent damage accumulation function.

***Transition to the erosion regime.*** We now focus on the transition between this regime of damage accumulation and the subsequent erosion of the material at large number of sliding cycles. Upon successive sliding, damage accumulates in the material, until it reaches a critical level sufficient to cause the detachment of matter from the outer surface. In our experimental conditions, this erosion process does not take place through the generation of particles, but rather through the formation of a sticky liquid-like "smearing" film, as shown in the photograph of Fig. 1D. Characterizing the rheological properties of this surface layer, we showed that it flows like a liquid with a high viscosity $\eta \sim 10^5$ Pa.s, suggesting that it is formed of highly degraded objects of submicronic sizes (see SI.S12). What controls exactly the transition between the two regimes of particulate formation and smearing is far from being clearly understood, as such transition might involves both mechanical considerations (stresses at the interface could be too weak to nucleate particulate formation through fracture (23), (39)) but also chemical ones. The formation of this degraded liquid film has indeed been shown to be inhibited in inert nitrogen atmosphere for filled industrial rubber (32). In the presence of oxygen, mechano-generated radicals react and are consumed by oxygen, leading to the progressive chemical degradation of the material (19), (33), (72), (73). In inert atmosphere, these mechano-generated radicals can recombine and the dominant mode for material removal from the surface becomes associated with particle generation through micro-fracturation.

We note finally that such distinction between particulate and smearing wear is also somewhat arbitrary: the observation of smearing wear in idealized laboratory conditions is not incompatible with wear through particle generation in real-world conditions. Indeed, in the presence of exogenous mineral particles (accounting for minerals inevitably present on road surfaces) the nature of the third-body generated by frictional wear is expected to change from a liquid-like material, to solid-like cohesive particles composed of gum mixed with mineral inclusions (20), (34), consistent with the structure of particles found on road (53). Smearing wear should thus not be discarded as an important and relevant wear mode for real-world conditions.

To analyze this erosion processes, we propose here a *local* evolution law for the damage field in the depth of the material $\phi(z, N)$, balancing the damage accumulation rate and the erosion rate and expressed as

$$\frac{d\phi(z)}{dN} = F(z, \phi) + \frac{\partial\phi(z)}{\partial z} \cdot \frac{dh}{dN} \qquad (Eq. 8)$$

Where $z$ in the distance to the (eroded) surface and $h$ is the absolute position of the surface. The condition for the detachment of material from the outer surface can be expressed through a local depercolation criterion associated with an intrinsic material damage threshold $\phi_{\mathrm{M}}$. The first term $F$ on the right-hand side characterizes an elementary damage increment function, dependent on the local value of the damage $\phi$, and on the distance $z$ from the surface through the stress field induced by the sliding asperities. The second term accounts for the effect of surface erosion, which leads to the convection of damage towards the surface, at a rate $v_{\mathrm{e}} = dh/dN$. In the first accumulation phase, $v_e = 0$ and the rescaling observed in Fig. 4B allows us to express the damage accumulation function as $F(z,\phi) \sim A_{\mathrm{r}}/A \cdot f(z,\phi)$ where $f$ is independent of the applied macroscopic pressure (Fig. 4B).

In the second regime of erosion, where steady-state damage has been reached (so that $d\phi/dN = 0$ and thus $d\phi/dz = F(z)/v_e$), we can express a depercolation criterion accounting for a critical level of damage at which materials will irreversibly leave the surface. We write this depercolation criterion as a condition on the cumulative damage experienced by a material point through the effect of the successive asperities, while it gradually reaches the surface via erosion. As this material point becomes closer to the surface, it is submitted to stresses and damage increments of growing amplitude. This balance can be expressed in mathematical terms as $\phi_{\mathrm{M}} = \int_0^{\infty} F\big(z(N)\big)dN = \frac{1}{v_{\mathrm{e}}}\int_0^{\infty} F(z)dz$, equating the cumulated damage (right-hand side) with the critical depercolation threshold $\phi_{\mathrm{M}}$.

This condition allows us to express the erosion speed as $v_{\mathrm{e}} = \frac{A_{\mathrm{r}}}{A}\int_0^{\infty} f(z)/\phi_{\mathrm{M}} dz$ and the subsurface damage $\phi(z) = \frac{A_{\mathrm{r}}}{A} \cdot \frac{1}{v_{\mathrm{e}}}\int_z^{\infty} f(z)dz$, giving a quantitative relation between the erosion properties of the material and the rate at which damage accumulates through successive asperity sliding.

The apparent independence of the detected damage on the applied normal force observed in the erosion regime in Fig. 4A is consistent with this picture. Coming back to our observation of an Archard-like law, we expect both the rate of damage accumulation and that of material erosion to approximately scale with the real area of contact $A_{\mathrm{r}}$, thus leading to a relative independence of the detected damage (in steady-state conditions) on pressure in the erosion regime (see SI S13). Equivalently, the prefactor $\frac{A_r}{A} \cdot \frac{1}{v_e}$ in our expression for subsurface damage $\phi(z)$ will show a weak dependence on normal force, in qualitative agreement with our data in Fig. 4.

Finally, we interpret the drop in detected surface damage $\bar{\Sigma}$ observed at the onset of the erosion regime as stemming from the lubricating effect of the liquid-like smeared third-body, which by decreasing stresses applied at the elastomer subsurface should subsequently lead to a reduction of the amount of accumulated damage. We highlight that such lubricating effect might not be specific to the presence of the smearing liquid as third-body generated particles are also expected to have a self-lubricating effect in the contact (25).

***Relevance for particulate wear.*** The picture proposed here of a continuous accumulation of damage in the subsurface, coupled to a critical depercolation threshold leading to erosion appears particularly well-suited to our experimental situation, whereby material leaves the surface under the form of a highly degraded liquid-like smearing layer.

This scenario and its associated physical picture is thus clearly distinct from recently proposed approaches for wear by particulate formation, which revisit early ideas proposed by Rabinovicz (74) and Grosch and Schallamach (20), and describe particulate formation through a balance between elastic energy at the contacting asperities and fracturation processes (23), (75). We note however, that the recent proposition to account for a fatigue-like process for the crack growth before particle detachment (23), (75) bears some similarity with the fatigue perspective we evidence for sub-surface damage accumulation.

Our physical picture highlighting the importance of the diffuse accumulation of subsurface damage could also be combined with the ideas of particulate detachment described above, to account for a final step of material removal for the outer surface by micro-fracturation (instead of the depercolation process proposed here, which could be seen as the limit where particulate size becomes very small). In this case, the presence of long-ranged diffuse damage in the subsurface would facilitate the final crack-growth process for particulate formation. The development of such combined models and their confrontation towards experimental data provides exciting perspectives for future studies.

Finally, we highlight again that this distinction between smearing wear and particulate wear is also somewhat arbitrary: when external mineral powders are added to the tribological test (to mimic mineral bodies inevitably present on road surfaces), smearing wear transition towards a mode of solid particle generation. These particles are composed of degraded polymer mixed with mineral inclusions, and this wear mode is expected to be a major one in engineering context (20), (34), (53).

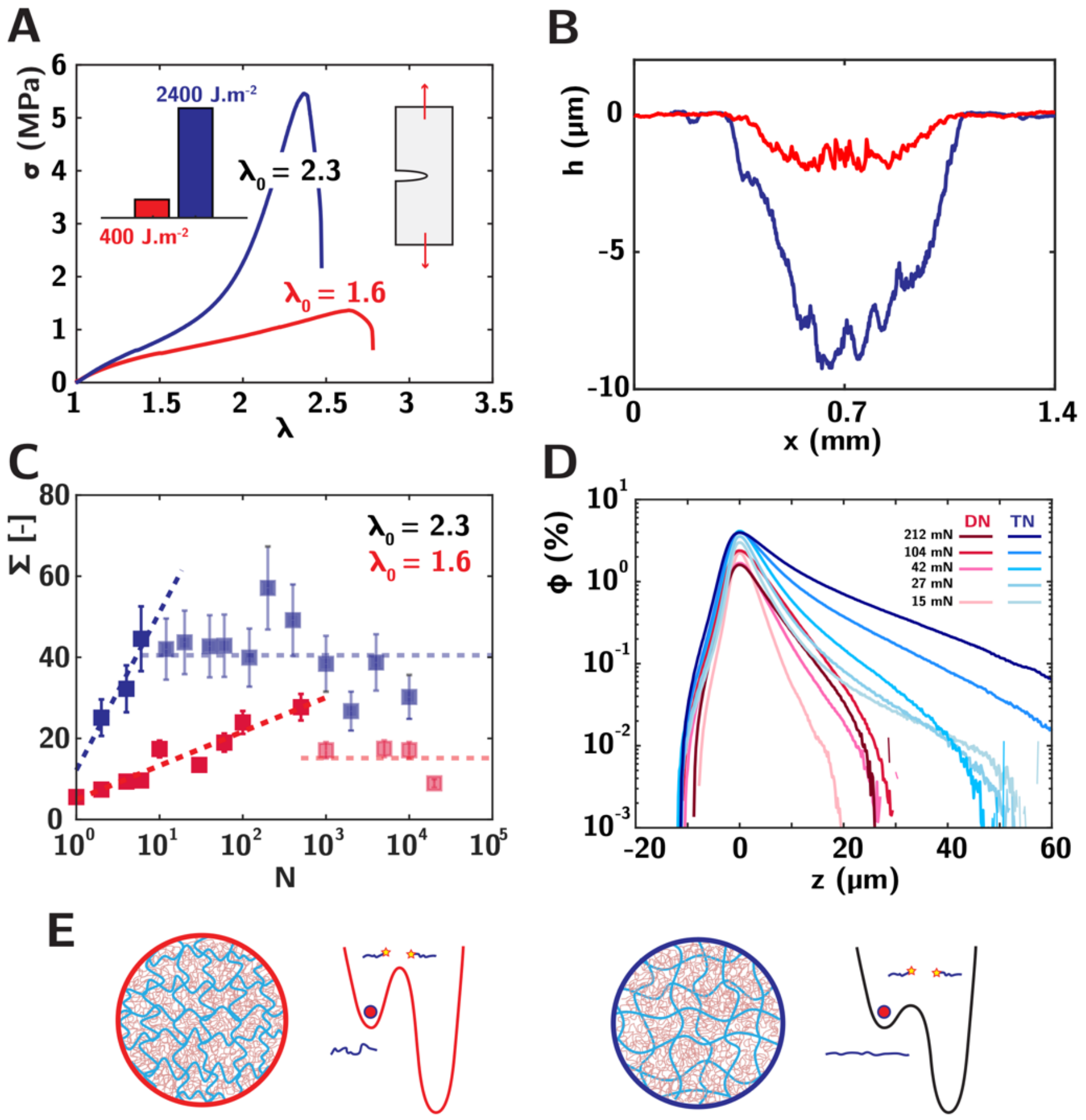

**Figure 5. Fracture/wear trade-off and the role of material architecture. (A)** Stress-deformation $\sigma(\lambda)$ on notched samples for double DN elastomer (red) and TN elastomer (blue) architectures. **(B)** Erosion profile for DN and TN elastomer at $N = 1000$ sliding cycles and $F_N = 42$ mN. **(C)** Evolution of the areal damage at $F_N = 15$ mN, for DN and TN elastomer. The dashed lines are logarithmic fits in the accumulation region. The horizontal dashed lines represent the average value in the erosion regime. **(D)** Evolution of spatial damage gradient for DN and TN elastomer networks in the erosion regime ($N = 5000$ and 2000 for DN and TN elastomer respectively) at increasing normal force (see legend). **(E)** Schematic of material architecture for DN elastomer (red) and TN elastomer (blue), and associated schematic bond scission potential.

***Material architecture reveals fracture/wear trade-off.*** A key question remains open, namely what are the molecular parameters that control the rate of damage accumulation and subsequent material erosion. To address this point, we compare in Fig. 5 two distinct materials characterized by a similar network architecture, with yet distinct levels of prestretch $\lambda_0$ of their filler network. The double network (DN) elastomer, whose frictional wear properties were presented in the core of this paper, is obtained through a single step of swelling and polymerization, leading to a moderate level of prestretch $\lambda_0^{\mathrm{DNE}} = 1.5$. We compare this material with a triple network (TN) elastomer obtained by the addition of a second swelling step leading to a higher prestretch $\lambda_0^{\mathrm{TNE}} = 2.3$ (SI.S1).

Fig. 5A shows representative stress-deformation curve for these two samples in uniaxial extension. The samples are prenotched, allowing to study their resistance to crack propagation and measure their fracture energy. The prestretch of the filler network bears a large impact on the associated mechanical properties. We observe in particular the occurrence of a pronounced strain hardening before crack propagation for the TN elastomer architecture, since the prestretched chains reach their finite extensibility at lower relative macroscopic deformation and the interpenetrated network architecture delays macroscopic crack propagation (76). Accordingly, the associated fracture energy measured with the Greensmiths's approximation (77) shows a marked dependence with network architecture, with fracture energies increasing from $\Gamma_{\mathrm{DN}} = 400\ \mathrm{J.m^{-2}}$ to $\Gamma_{\mathrm{TN}} = 2400\ \mathrm{J.m^{-2}}$ (inset, Fig. 5A) (46).

Remarkably, a strictly opposite behavior is observed when comparing the fatigue wear resistance of both materials. We report in Fig. 5B the surface wear profile for DN (red) and TN elastomer (blue) for a given condition of $F_{\mathrm{N}} = 42$ mN and $N = 1000$. The deeper profile is associated with a clearly poorer wear resistance for the TN elastomer. This trend was confirmed for various conditions of cycle number and normal force, from which we extract a wear probability coefficient $k$, relating the worn volume to the volume of sled interface $V \approx k.A_{\mathrm{r}}.l$ and which increases from $k_{\mathrm{DN}} = 0.9.10^{-6}$ to $k_{\mathrm{TN}} = 6.10^{-6}$ for respectively DN and TN elastomer networks (See SI.S6).

This peculiar trade-off between resistance to crack propagation and frictional wear resistance calls for explanations. To rationalize these trends, we resort in Fig. 5C to our mechanosensitive approach and probe how TN and DN elastomer compare in terms of damage accumulation. We compare here the evolution of damage - expressed in terms of number of broken molecular layers $\bar{\Sigma}$ - in DN and TN elastomer at a fixed normal force $F_{\mathrm{N}} = 15$ mN. As shown in this figure, we recover a logarithmic-like regime of damage accumulation following friction cycles for the TN elastomer (blue dotted line) yet with a much higher rate than for the DN elastomer, associated with a pre-logarithmic factor increasing from $\approx$ 4 to 17. Coming back to our molecular picture of damage accumulation evidenced in Fig. 4, this higher wear rate suggests that the network of the TN elastomer is characterized by much lower energy barriers for bond scission. This increased sensitivity to frictional stresses is also clearly apparent in Fig. 5D, where we focus on the spatial distribution of damage in the erosion regime, compared for the DN and TN elastomer at various normal loads. The TN elastomer is characterized here by the occurrence of spatial profiles which extend systematically to larger distances below the elastomer surface. Subsurface damage in the TN elastomer material is also characterized by a larger sensitivity to the normal load, with damage extension increasing with $F_{\mathrm{N}}$. While the dominant effect of the large normal forces will be to increase the number of contacting asperities thus increasing the rate of damage accumulation, the maximal stresses borne by the largest asperities should also increase slightly, leading to the propagation of a stress gradient deeper in the material. The larger sensitivity to stress fluctuation evidenced for the TN elastomer might thus lead to the extension of molecular damage deeper in the material in the conditions of high normal forces.

The unexpected trade-off between fatigue and fracture behavior is puzzling - yet this peculiar result can be illuminated with our mechanosensitive approach combined with our fine control of the network architecture. As schematically represented in Fig. 5C, the higher pre-stretch for the TN elastomer architecture dilutes the filler chains in the matrix network and brings them closer to their limiting extensibility and tipping point for failure. When probing the network resistance in terms of tensile properties this results in the so-called "sacrificial bond concept", whereby the TN elastomer architecture allows for the delocalization of stresses and bond scission (46) leading to a spatially extended damage zone which delays the nucleation and propagation of cracks (76), (78). In particular, this protective effect associated with damage delocalization is operative as crack propagation is a one-shot event associated to very large strains at the crack tip. On the contrary, fatigue wear resistance is set by the resistance of the material to damage accumulation under a *high number* of *low-intensity* solicitations. In this context, the presence of weaker sacrificial links as present in the TN elastomer appear detrimental as they would lead to a dramatic increase of the rate of damage accumulation (as evidenced in Fig. 5B-E) and thus faster macroscopic wear.

The occurrence of such fracture/wear trade-off on this class of multiple network materials is unexpected *per se*. Such trade-off has seldom been reported in the literature (79), where it has been attributed to geometrical effects and surface reconstruction for blade abrasion, rather than molecular material parameters in terms of stress redistribution and damage accumulation. Probing the possible generalization of the concepts presented here for multiple network elastomers to more complex and industrially relevant materials such as filled rubber would thus offer exciting perspectives for future work.

***Conclusion.*** Despite its major importance, current mechanistic understanding of elastomer wear remains limited, due to the difficulty to assess the local damage field in the material following mild frictional events. We proposed here a novel approach using damage-sensitive mechanochemical probes, revealing damage by chain scission following frictional sliding on ~~in~~ model elastomer materials based on the multiple network architecture. Using this mechanosensitive approach, we elucidated the mechanisms underlying wear of our elastomeric materials, revealing that mild wear does not originate from crack propagation but rather from the continuous accumulation of subsurface damage, ultimately leading to material erosion through the generation of liquid-like smearing film. This damage is induced by the rough sliding asperities, extending well below the material surface and accumulating in a spatially heterogeneous manner through discrete micro-slippage events at the asperities. The damage accumulation process further follows a slow, logarithmic-like growth, indicative of stress-activated scission events within a heterogeneous elastic energy landscape. Our findings highlight the probabilistic nature of this fatigue-like damage accumulation mechanism, allowing us to formulate the wear rate as an integral of the cumulative damage over successive asperity sliding events. The slow yet steady accumulation of damage near the surface, ultimately couples to a depercolation process and leads to material removal from the interface. Finally, by tuning the molecular architecture of our materials, we uncovered an antagonistic relationship between fracture resistance and wear resilience, governed by the material's sensitivity to stress fluctuations. Beyond providing a foundation for future advancements in tribology and material science, the implications of these findings are both industrially environmentally relevant. Our work underscores the critical role of previously unobserved subsurface damage in controlling elastomeric wear, offering novel insights that should guide further physics-based approaches towards better wear-resilient materials.

## Materials and methods

### *Materials*

The materials presented here were synthesized from monomers. The synthesis of double and triple networks with the pi-extended anthracene mechanophore was previously described in (46) and is briefly recalled here.

Ethyl acrylate monomer (EA, Sigma Aldrich, 99%) was purified prior to the synthesis by chromatography to remove the MEHQ added as polymerization inhibitor. 1,4-butanediol di-acrylate (BDA, Sigma) and 2-hydroxy-2-methylpropiophenone (HMP, Sigma) were used as received. The synthesis was carried out in a glovebox (Mbraun Unilab) under nitrogen atmosphere, to avoid side reactions with the oxygen from the air during polymerization. The labelling of the network was done using a mechano-fluorescent probe, the Diels-Alder diacrylate cross-linker (DACL) (see (47) for details on the DACL synthesis).

### *Filler network synthesis and incorporation of mechanophore*

BDA was used as a conventional cross-linker (0.47 mol%, with respect to monomer concentration) and HMP (1.16 mol%) as a radical photo initiator. The DACL mechanophore was covalently incorporated as a cross-linker (0.03 mol%) in combination with BDA, for a total cross-linker concentration of 0.5 mol%. 5.6% of the cross-linkers were mechanophore cross-linkers. The final concentration of mechanophore in the network was 2.55 mol.m-3, this quantity was sufficient to allow for fluorescence quantification while leaving the network mechanical properties unaffected.

EA monomer, BDA, DACL and HMP were mixed together and introduced into a mold made of two glass plates, coated with a transparent PET film, to reduce adhesion to the wall during polymerization. The thickness of the sample was controlled using a silicon spacer (0.5 mm). The filled mold was placed in front of a UV lamp ($\lambda = 365$ nm, Power $< 10$ µW.cm-2) for 2 hours to polymerize and crosslink. The residual monomer was then left to evacuate under vacuum at r.t. overnight. Weight loss during drying was about 0.1 %, indicating a good conversion during polymerization.

### *Double and triple network synthesis*

The preparation of double and triple networks required successive steps of swelling of the filler network in monomer, followed by polymerization. EA monomer (18 mL), BDA (0.01 mol%) and HMP (0.01 mol%) were mixed together. A 40x20mm single network was placed into this solution and left to swell for 2 hours. The swollen sample was then placed in a mold made of 2 glass plates coated with PET sheets. The rest of the synthesis (UV, drying) is the same as described above.

As described in (46), the double network elastomer (DNE) obtained was characterized by the fraction of filler network ($\phi_{SN}$) within the final network, and equivalently, by the level of isotropic pre-stretching of the network ($\lambda_0$), due to the swelling phase of the single network. These two quantities could be calculated from the mass of the filler (single network elastomer SNE) before swelling ($m_{SN}$) and the mass after polymerization ($m_{tot}$), using:

$$\phi_{SN} = \frac{m_{SN}}{m_{tot}} \text{ , } \lambda_0 = (\phi_{SN})^{-1/3}$$

The swelling and polymerization steps were repeated a second time to obtain triple network elastomer (TNE), in which the swelling of the first network leads to a higher level of pre-stretching. More details on the final pre-stretch and characteristics of these networks are given in SI.1.

***Friction setup***
A homemade friction setup was used to perform wear tests on elastomers. A plano convex glass lens, with a radius of curvature of 5.19 mm (SLB-08-10P from OptoSigma) was mounted on a metal bar free to rotate in the transverse direction to the sliding direction, allowing for permanent contact between the glass sphere and the material. A dead-load on top of the lens was used to impose a constant normal force. The sample was set on a platform fixed to a precision linear stage (M404.6PD from PI) performing reciprocating motion. Stage speed and displacement were monitored via a motor controller (C-863 Mercury Controller from PI) through a script on PI-Mikro Move. The measurement of the tangential friction force was acquired from the deflection of two thin metal springs mounted on each side of the moving platform. Deflection was measured using a capacitive controller (capaNCDT 6110 from Micro-Epsilon) and recorded via a LabView script. A schematic representation of the setup is given in SI. 2.

The effective stiffness of the bimetallic strips was estimated to $k \approx 5 \cdot 10^4\ \mathrm{N.m^{-1}}$. For all the wear tests, the velocity of the moving platform was set to $2\ \mathrm{mm.s^{-1}}$. The platform was moving back and forth over a distance of 7 mm. We note 1 cycle for each one-way motion of the platform.

***Optical profilometry***
Macroscopic wear assessment was performed with a Wyko NT9100 optical profilometer, using the Vertical Scanning Interferometer (VSI) mode with a 5x objective, and a zoom factor of 0.55. Each acquisition was conducted over a 2.28*1.71 mm zone (640*480 pixels, with a pixel size of 3.56 µm) large enough in the y-direction to cover the wear mark and part of the unworn material. For some acquisitions involving larger wear marks, stitching of multiple images was conducted to cover a larger zone around the wear mark.

***Confocal imaging***
The fluorescent signal of the mechanophores was measured using a Leica TCS SP8 CSU confocal microscope. In most of our acquisitions, we used a 40x inverted oil immersion objective, with a zoom factor 0.75. Image size was 388.26 x 388.26 µm (512 x 512 pixels). Pixel size was 758 x 758 nm. Due to the short working distance (0.17 mm), the sample was placed on a cover slip and a few drops of glycerol ($n = 1.473$) were placed between the worn surface of the sample ($n = 1.464$) and the glass cover slip to avoid index mismatch due to air (80). The excitation wavelength was 405 nm, and fluorescence emission was collected between 450 and 550 nm. To ensure a z-resolution as good as possible, the pinhole was closed to its smallest value. With this strategy, a nominal optical section of 0.59 µm was reached (indicated by the Leica software). Stacks of images were acquired (z-stacks) and the $\Delta z$-step between two consecutive pictures was 0.20 µm. The global acquisition range in z was chosen so that the average intensity on the first image – outside the material and above the fluorescent surface, and last images – below the activated surface, was as low as possible (the total z-range was typically $\sim 45$ µm). Since the 40x objective could not cover the entire width of the wear mark, in our analysis, the acquisition was done in the center of the wear mark.

**Acknowledgments**
The authors thank Paul-Aymé Toulemonde for contributions at the early stage of the project. J.Comtet and O.T. thank Antoine Chateauminois for interesting discussions and feedback. The authors are indebted to Matthias Nebouy and Gabriel Sanoja for discussions and feedback on the manuscript. This work benefited from the technical contribution of the joint service unit CNRS UAR 3750 and was supported by UAR 2063 Workshop for Advanced Research Support.

***Funding:*** We gratefully acknowledge the financial support of Michelin and of the Association Nationale de la Recherche et de la Technologie.

***Author contributions:*** J.Comtet and C.C. conceived the project and designed research. O.T. and C.T.B. synthesized samples and performed experiments and data analysis with inputs from J.Comtet, C.C., M.C. and J.Caillard. O.T., J.Caillard, M.C., C.C. and J.Comtet interpreted data. J.Comtet and J.Caillard developed the modelisation approach. J.Comtet. and O.T. wrote the manuscript with inputs from all authors.

***Competing interests:*** All authors declare that they have no competing interests.

***Data, Code, and Materials Availability:*** All data needed to evaluate and reproduce the results in the paper are present in the paper and/or the Supplementary Materials. This study did not generate new materials.

**References.**